\documentclass[aps,prl,twocolumn,showpacs,showkey,square,numbers,amssymb,amsmath,nobibnotes]{revtex4-1}
\usepackage{bm}
\usepackage{times,float}
\usepackage{graphicx}
\usepackage[usenames,dvipsnames,svgnames]{xcolor}
\usepackage{hyperref}
\hypersetup{colorlinks=true, linkcolor=NavyBlue, citecolor=PineGreen,urlcolor=cyan}

\newcommand{\bracket}[1]{\left\langle #1\right\rangle}

\begin{document}
\title{Theory for the conditioned spectral density of non-invariant random matrices}
\author{Isaac P\'erez Castillo}
\address{Department of Quantum Physics and Photonics, Institute of Physics, UNAM, P.O. Box 20-364, 01000 Mexico City, Mexico}
\address{London Mathematical Laboratory, 14 Buckingham Street, London WC2N 6DF, United Kingdom}
\author{Fernando L. Metz}
\address{Institute of Physics, Federal University of Rio Grande do Sul, 91501-970 Porto Alegre, Brazil}
\address{Physics Department, Federal University of Santa Maria, 97105-900 Santa Maria, Brazil}
\address{London Mathematical Laboratory, 14 Buckingham Street, London WC2N 6DF, United Kingdom}

\begin{abstract}
  We develop a theoretical approach to compute the conditioned spectral density of $N \times N$ non-invariant
  random matrices   in the limit $N \rightarrow \infty$.  This large deviation observable, defined as the eigenvalue  distribution conditioned to have a fixed fraction $k$ of eigenvalues smaller than $x \in \mathbb{R}$, provides the spectrum of random matrix samples that deviate atypically from the average behavior.   We apply our theory to sparse random matrices and unveil  strikingly new and generic properties, namely: (i) their conditioned spectral density has compact support; (ii) it does not experience any abrupt transition for $k$ around its typical value; (iii) its eigenvalues  do not accumulate at $x$. Moreover, our work points towards other types of transitions in the conditioned spectral density for values of $k$ away from its typical value. These properties follow from the weak or absent eigenvalue repulsion in sparse ensembles and they are in sharp contrast to  those displayed by  classic or  rotationally invariant random matrices.  The exactness  of our theoretical findings are confirmed through numerical diagonalization of finite random matrices. 
\end{abstract}
\maketitle

Ensembles of random matrices offer the simplest non-trivial mathematical laboratory to study the statistical properties
of rare events in complex systems. Thanks to the universal properties of their eigenvalue statistics, random matrices find applications in areas as diverse as nuclear physics, quantum chaos, finance, complex networks, and string theory \cite{BookMehta,akemannBook}.

The study of classical disordered systems is an emblematic example where random matrix theory is very useful. Large many-particle systems with quenched or self-induced disorder are described by a free-energy or an energy function \cite{wales}, respectively. The structure of such functions in the configuration space is usually highly nontrivial, owing to the enormous number of  stationary points  (saddle-points, minima and maxima), whose impact on the dynamics of the system is decisive \cite{wales,Kurchan,Parisi2002}. The local stability  around an extremum is determined by the Hessian matrix of the energy second derivatives, with its positive (negative) eigenvalues measuring the curvature of the stable (unstable) directions. Unfortunately, deriving the distribution of the Hessian is a hopeless task, due to its explicit dependency on the configurations, and a fruitful strategy consists in replacing the Hessian by a suitable random matrix \cite{Cavagna2000}. In this approach, the distribution of the number of eigenvalues below a certain threshold, the so-called {\it index}, is exactly computed \cite{Cavagna2000,Majumdar2009,Isaac2016}. This effective model for the Hessian has led to valuable insights not only in disordered systems \cite{Cavagna2000}, but also in  string theory \cite{String1,String2} and quantum cosmology  \cite{quantumcosm1,quantumcosm2}.

Even though the index distribution characterizes the fluctuations of the stability throughout the energy landscape, one may be interested, for instance, in the curvature distribution around specific saddle-points with a given number of stable directions. In order to access such refined information, one has to compute the {\it conditioned spectral density} (CSD), namely the Hessian eigenvalue distribution constrained to have a {\it fixed} fraction of positive eigenvalues. As the typical fluctuations of this fraction vanish by increasing the system size, the CSD yields the spectrum of constrained rare samples, strongly deviating from the typical case.

The CSD is a crucial observable in random matrix theory. This quantity arises as a byproduct of the Coulomb fluid method \cite{Dyson3}, recently adapted to obtain the index distribution in the case of {\it invariant random matrix ensembles} \cite{Majumdar2009,Vivo,Majumdar3,Marino}, which include Gaussian \cite{Majumdar2009,castillo2014spectral,castillo2016large}, Wishart \cite{Vivo,melo2015unified}, and Cauchy random matrices \cite{Majumdar3}. In this context, the effect of constraining the eigenvalues in different regions is modeled as having one or more confining walls. For invariant random matrices, the CSD exhibits generic features, resulting from the repulsive Coulomb interaction between eigenvalues, namely: i) it is an asymmetric function with a non-compact support; (ii) it undergoes an abrupt transition as the number of constrained eigenvalues crosses its typical value \cite{Marino}; iii) the eigenvalues accumulate at the walls.

A natural question is whether these salient and somewhat universal features persist in the case of {\it non-invariant} random matrices, for which the joint distribution of eigenvalues is generally unknown and, consequently, the Coulomb fluid method is inapplicable. One of the most interesting classes of non-invariant random matrices is that of  {\it sparse random matrices} (SRMs), whose main defining property is the presence of a large amount of zero entries. Since SRMs encode the topology of spin models on tree-like random graphs, they find applications in several fields, including spin-glasses, error-correcting codes, optimization problems, and complex networks (see \cite{MezardBook,Barrat} and references therein). The spectrum of SRMs  is richer than their invariant counterparts \cite{Perez2008,Perez2009,Perez2010}, with the existence of regions containing localized eigenvectors \cite{Evangelou1, Kuhn2008, Metz2010,Slanina2012}, in which eigenvalue repulsion is weak or absent. It is unclear how this feature affects the properties of the CSD.

In this Letter we present a novel theoretical approach that allows to compute the CSD of non-invariant random matrices in the limit $N \rightarrow \infty$. We apply our technique to two paradigmatic ensembles of SRMs: the adjacency matrix of Erd\"os-R\'enyi random graphs and sparse Wishart random matrices. Our analysis shows that new universal properties emerge in the case of SRMs: i) the conditioned spectral density displays a compact support; ii) it does not exhibit any transition when the constrained number of eigenvalues approaches its typical value; iii) in the limit $N \rightarrow \infty$, there is no accumulation of eigenvalues at the wall. All these properties seem to follow from the absence of eigenvalue repulsion, in strike contrast with the behavior of traditional invariant ensembles. The theoretical results are fully confirmed by numerical diagonalization of finite random matrices.

We consider an ensemble of $N \times N$ symmetric random matrices $\bm{M}$ with eigenvalues $\{\lambda_i \}_{i=1,\dots,N}$. The number of eigenvalues smaller than a threshold $x \in \mathbb{R}$ is given by  $\mathcal{I}_N(x)=\sum_{i=1}^N\Theta(x-\lambda_i)$, where $\Theta(x)$ is the Heaviside step function. The conditioned spectral density is defined as
\begin{equation}
  \rho_x(\lambda|k)=\lim_{N\to\infty} \frac{N^{-1} \bracket{\sum_{i=1}^N\delta(\lambda-\lambda_i)\delta\left[kN - \mathcal{I}_N(x) \right]}}{\bracket{\delta\left[kN - \mathcal{I}_N(x)\right]   }  } ,
  \label{ss1}
\end{equation}
where $\delta$ is the Dirac delta and $\bracket{\dots}$ denotes the expectation over the random matrix ensemble. Equation (\ref{ss1}) represents the conditional probability density for the eigenvalues between $\lambda$
and $\lambda + d \lambda$, provided there is precisely $k N$ eigenvalues smaller than $x$. Notice that we are imposing a hard constraint in the random matrix ensemble by choosing those samples for which $\mathcal{I}_N(x) = kN$, having $x \in \mathbb{R}$ and $0 \leq k \leq 1$ as adjustable parameters of our theory.

Let us sketch our theoretical approach to evaluate the CSD for arbitrary random matrix ensembles. Here we state only the main results, while all details of our technique are discussed in the Supplemental Information \footnote{See details in the Supplemental Information, which also contains  the Ref. \cite{coolen2016replica}}. Using the standard version of the replica method \cite{BookParisi}, combined with a representation of the index  $\mathcal{I}_N(x)$ in terms of complex logarithms \cite{Cavagna2000,Metz2015}, one rewrites the CSD as 
\begin{eqnarray}
&\rho_x&(\lambda|k)=- \frac{2}{\pi} \lim_{\eta\to{0}^+}\lim_{\epsilon\to 0^{+}}\lim_{N\to\infty}\lim_{n\to 0} \label{eq:2}  \\
&\times& \frac{1}{N n}\frac{\partial}{\partial \lambda_{\eta}}
\text{Im} \left[ \int dy \,  \mathcal{P}^{(N)}(y,x_\epsilon) \ln \left(\frac{Q^{(N)}_n(y,\lambda_\eta,x_\epsilon)}{  Q^{(N)}_{0}(y,x_\epsilon)   }\right)\nonumber
  \right],
\end{eqnarray}
where the weight $\mathcal{P}^{(N)}(y,x_\epsilon)$ is given by
\begin{equation}
  \mathcal{P}^{(N)}(y,x_\epsilon) =  \frac{\exp{\left(N\left[yk -\mathcal{F}^{(N)}(y,x_\epsilon)\right]\right)}}
          {\int dy \exp{\left( N\left[yk -\mathcal{F}^{(N)}(y,x_\epsilon)\right]\right) } },
\end{equation}
with
\begin{align}
  Q^{(N)}_{n}(y,\lambda_\eta,x_\epsilon) &=  \Big\langle \left[\overline{Z(x_\epsilon)}\right]^{\frac{i y}{\pi}}
    \left[Z(x_\epsilon)\right]^{ -\frac{i y}{\pi}}    [Z(\lambda_\eta)]^n \Big\rangle, \label{ssw1} \\
       \mathcal{F}^{(N)}(y,x_\epsilon) &=\frac{y}{2}-\frac{1}{N} \ln  Q^{(N)}_{0}(y,x_\epsilon)  . \label{ssw3}
\end{align}
The symbol $\overline{(\dots)}$ stands for complex conjugation. The function $Z$ explicitly depends on the random matrix $\bm{M}$ and it is defined generically as follows
\begin{eqnarray}
  Z(a_\mu)&=&\int_{-\infty}^{\infty} \left(  \prod_{i=1}^N dy_i \right) \exp\left[-\frac{i}{2} \bm{y}^T.(a_\mu\mathbb{I}-\bm{M}) \bm{y} \right],  \nonumber \\
  a_\mu &=& a - i \mu,  \qquad \bm{y}^T = (y_1,\dots,y_N),
\end{eqnarray}
with $\mathbb{I}$ denoting the $N \times N$ identity matrix. Equation (\ref{ssw1}) is a direct consequence of introducing mathematical identities to represent $\mathcal{I}_N(x)$ and the Dirac delta \cite{Metz2015,Metz2016,Perez2018}, while Eq. (\ref{ssw3}) determines the cumulant generating function that controls the fraction of eigenvalues smaller than $x$ \cite{Note1}.

The derivation of Eq. (\ref{eq:2}) is a fundamental step of our analytic approach, since it recasts the calculation of $\rho_x(\lambda|k)$ in terms of the solution of an extremization problem. In fact, let us assume the limit $N \rightarrow \infty$ can be  performed  before taking $n \rightarrow 0$ in Eq. \eqref{eq:2}. This is a common and harmless procedure in the study of disordered systems \cite{BookParisi}, which enables us to evaluate formally the integral over $y$ in Eq. \eqref{eq:2} using the saddle-point method
\begin{eqnarray}
\rho_x(\lambda  |  k)&=& - \lim_{\eta\to{0}^+}\lim_{\epsilon\to0^{+}}\lim_{n\to0}\nonumber\\
&&\hspace{-1cm}\times\frac{2}{\pi n} \frac{\partial}{\partial \lambda_{\eta}} \text{Im}   \left(\lim_{N\to\infty}\frac{1}{N}
\ln \left[Q^{(N)}_n(y_\star,\lambda_\eta,x_\epsilon)\right]\right),
\label{ssap}
\end{eqnarray}
where $y_\star$ is obtained from the stationary condition
\begin{equation}
  k=\frac{\partial \mathcal{F}(y,x_\epsilon)}{\partial y}\Big|_{y=y_\star},
  \label{aaqw}
\end{equation}
with $\mathcal{F}(y,x_\epsilon) \equiv \lim_{N \rightarrow \infty}  \mathcal{F}^{(N)}(y,x_\epsilon)$. This concludes the essence of our analytic approach:  according to Eq. (\ref{ssap}), the computation of $\rho_x(\lambda  |  k)$ boils down to being able to calculate $Q^{(N)}_{n}(y,\lambda_\eta,x_\epsilon)$ for $N \rightarrow \infty$. Note that, up to this point, our technique is completely general, in the sense we have not done any assumption regarding the random matrix ensemble.

In principle, once the random matrix ensemble is specified, any suitable analytic tool can be used  to evaluate the ensemble average and the large-$N$ behavior of $Q^{(N)}_n$, yet the presence of imaginary and real-valued exponents in Eq. (\ref{ssw1}) is a serious issue. We surmount this obstacle by following the replica approach as discussed in \cite{Metz2015,Metz2016,Perez2018}. At first, these exponents are regarded as integer positive numbers, which allows to calculate the ensemble average and extract the large-$N$ behavior of  $Q^{(N)}_n$. The last step is the analytic continuation of the exponents $n_{\pm}$ and $n$ back to, respectively, their original complex and real values, which is only feasible if one makes an {\it ansatz} regarding the functional form of the order-parameter emerging in the effective replica theory. Here we present analytic expressions for the CSD under the so-called replica-symmetric assumption for the order-parameter \cite{BookParisi,Metz2016,Perez2018}.

We derive explicit results for two prototypical ensembles of SRMs: the adjacency matrix of weighted Erd\"os-R\'enyi (ER) random graphs \cite{Bollobas}, and sparse Wishart random matrices \cite{Perez2008}. In the first case, the matrix entries are defined as $M_{ij} = c_{ij} J_{ij}$, where the random variables $\{ c_{ij} \}$ are independently and identically drawn from 
\begin{equation}
  P_c(c_{ij}) = \frac{c}{N} \delta_{c_{ij},1} + \left( 1 - \frac{c}{N}  \right) \delta_{c_{ij},0} , \label{ffg}  
\end{equation}
with $c_{ii} = 0$ and $c_{ij} = c_{ji}$. The parameter $c$ is the average connectivity of the corresponding ensemble of ER random graphs \cite{Bollobas}. The variables  $\{ J_{ij} \}$ set the values of the nonzero entries of $\bm{M}$ and they are drawn independently from a  distribution $P_J(J)$, whose analytic form does not need to be specified at this point.
In the sparse Wishart ensemble, $\bm{M}$ is an $N \times N$ sparse covariance matrix $\bm{M}=c^{-1} \bm{\xi}\bm{\xi}^T$ built from the $N\times P$ rectangular random matrix $\bm{\xi}$. Its entries $\{ \xi_{i\mu} \}$, with $i=1,\ldots,N$ and $\mu=1,\ldots,P$, are independently and identically drawn from the distribution $P_c(\xi_{i\mu} )$, defined in  Eq. (\ref{ffg}).
From a random graph viewpoint, $\bm{\xi}$ is associated to a bipartite random graph, in which the $\mu$-nodes have average degree $c$, while the $i$-nodes have average degree $\alpha c$, with $\alpha=P/N$ \cite{Perez2008}. In both random matrix ensembles, $\bm{M}$ has a  sparse structure as $c$ remains of $O(1)$ in the limit $N \rightarrow \infty$. 

Although we show explicit results for both ensembles introduced above, the final equations determining
the CSD are presented only for the ensemble of ER random graphs (the analogous formulas for the sparse Wishart ensemble are shown in \cite{Note1}). Thus, we have obtained the following expression for the CSD
\begin{equation}
  \rho_x(\lambda|k)= \frac{1}{\pi} \lim_{\eta\to{0}^+ }\lim_{\epsilon \to 0^{+}}  \int_{-\infty}^{\infty}
  d\Gamma w_{y_\star}(\Gamma)\ \text{Im} \Gamma\,,
\label{eq:4}
\end{equation}
where $ d\Gamma \equiv d\text{Im}\Gamma d\text{Re}\Gamma$,  and $w_{y_\star}(\Gamma) $ is the joint probability
density of the real and imaginary parts of $\Gamma$. For the sake of simplicity, we have omitted the
dependence of $w_{y_\star}(\Gamma)$ with respect to $x_\epsilon$. The function $w_{y_\star}(\Gamma) $ follows from
\begin{equation}
  w_{y_\star}(\Gamma)= \int_{-\infty}^{\infty} d\Delta \, w_{y_\star}(\Delta,\Gamma), \label{laqp}
  \quad d\Delta \equiv d\text{Im}\Delta d\text{Re}\Delta,
  \end{equation}
where $w_{y_\star}(\Delta,\Gamma)$ obeys the self-consistency equation in the case of ER random graphs
\begin{eqnarray}
  w_{y_\star}(&\Delta&,\Gamma)= \frac{1}{\mathcal{N}} \sum_{\ell=0}^\infty\frac{e^{-c} c^\ell}
  {\ell!}\int_{-\infty}^{\infty} \left[\prod_{k=1}^\ell d\Delta_k  d\Gamma_k~w_{y_\star}(\Delta_k,\Gamma_k) \right] \nonumber \\
&\times& \Bigg\langle \mathcal{W}_{y_\star}[\{J_k,\Delta_k\}_{k=1}^\ell] \,
\delta\left(\Delta-\frac{1}{x_\epsilon-\sum_{k=1}^\ell J_k^2\Delta_k}\right)  \nonumber \\
&\times&
\delta\left(
\Gamma-\frac{1}{\lambda_\eta-\sum_{k=1}^\ell J_k^2\Gamma_k}\right)
\Bigg\rangle_{J_1,\dots,J_\ell}
\,,
\label{eq:sce1}
\end{eqnarray}
with the weight
\begin{equation}
  \mathcal{W}_{y_\star}[\{J_k,\Delta_k\}_{k=1}^\ell]\equiv 
    \left(\frac{x_\epsilon-\sum_{k=1}^\ell J_k^2\Delta_k}{\overline{x_\epsilon}-\sum_{k=1}^\ell J_k^2\overline{\Delta}_k} \right)^{ \frac{i y_\star}{2\pi} } .
\end{equation}
The brackets $\langle \dots \rangle_{J_1,\dots,J_\ell}$ represent the average over $J_1,\dots,J_\ell$ with
the distribution $P_J(J_k)$ ($k=1,\dots,\ell$), while $\mathcal{N}$ is the normalization factor ensuring $\int_{-\infty}^{\infty} d\Delta d\Gamma
w_{y_\star}(\Delta,\Gamma)=1$. The
value of $y_\star$ follows from the solution of Eq. (\ref{aaqw}), where the cumulant generating function for ER random graphs reads
\begin{eqnarray}
\mathcal{F}&(y,x_{\epsilon})&=y+\frac{c}{2} \Bigg\langle \int_{-\infty}^{\infty} d \Delta d \Delta^{\prime} w_y(\Delta) w_y(\Delta^{\prime}) \nonumber \\
&\times&\left[ \left(\frac{1-J^2 \Delta \Delta^{\prime}}{1-J^2 \overline{\Delta\Delta^{\prime}}}\right)^{ \frac{iy}{2\pi} } - 1\right]
\Bigg\rangle_{J}  \nonumber \\
&-&\ln \Bigg\langle \sum_{\ell =  0}^\infty \frac{e^{-c} c^\ell}{\ell!} \int_{-\infty}^{\infty}  \left[\prod_{k=1}^\ell d \Delta_k w_y(\Delta_k)\right]
\nonumber  \\
&\times&
\mathcal{W}_{y}[\{J_k,\Delta_k\}_{k=1}^\ell] \Bigg\rangle_{J_1,\dots,J_\ell}. \nonumber
\end{eqnarray}
Given a position of the threshold $x\in\mathbb{R}$, we determine $y_\star$ corresponding to an imposed fraction $k$ of eigenvalues smaller than $x$. This is done  by solving iteratively the fixed-point Eq. (\ref{aaqw}) for  $y_\star$ using Newton's method, as explained in  \cite{Note1}.

Let $k_{\text{typ}}(x)$ be the typical fraction of eigenvalues smaller than $x$, obtained by integrating the unconditioned spectral density from $-\infty$ to $x$ \cite{Perez2008}. We may have an excess or a defect of eigenvalues to the left of $x$ depending if we choose $k>k_{\text{typ}}(x)$ or $k<k_{\text{typ}}(x)$, respectively. We want to understand  how the eigenvalues accommodate themselves as $k$ changes and compare the results with those obtained for
invariant random matrices \cite{Majumdar2009,Vivo,Majumdar3}.

\begin{figure}
  \includegraphics[width=8.5cm,height=6cm]{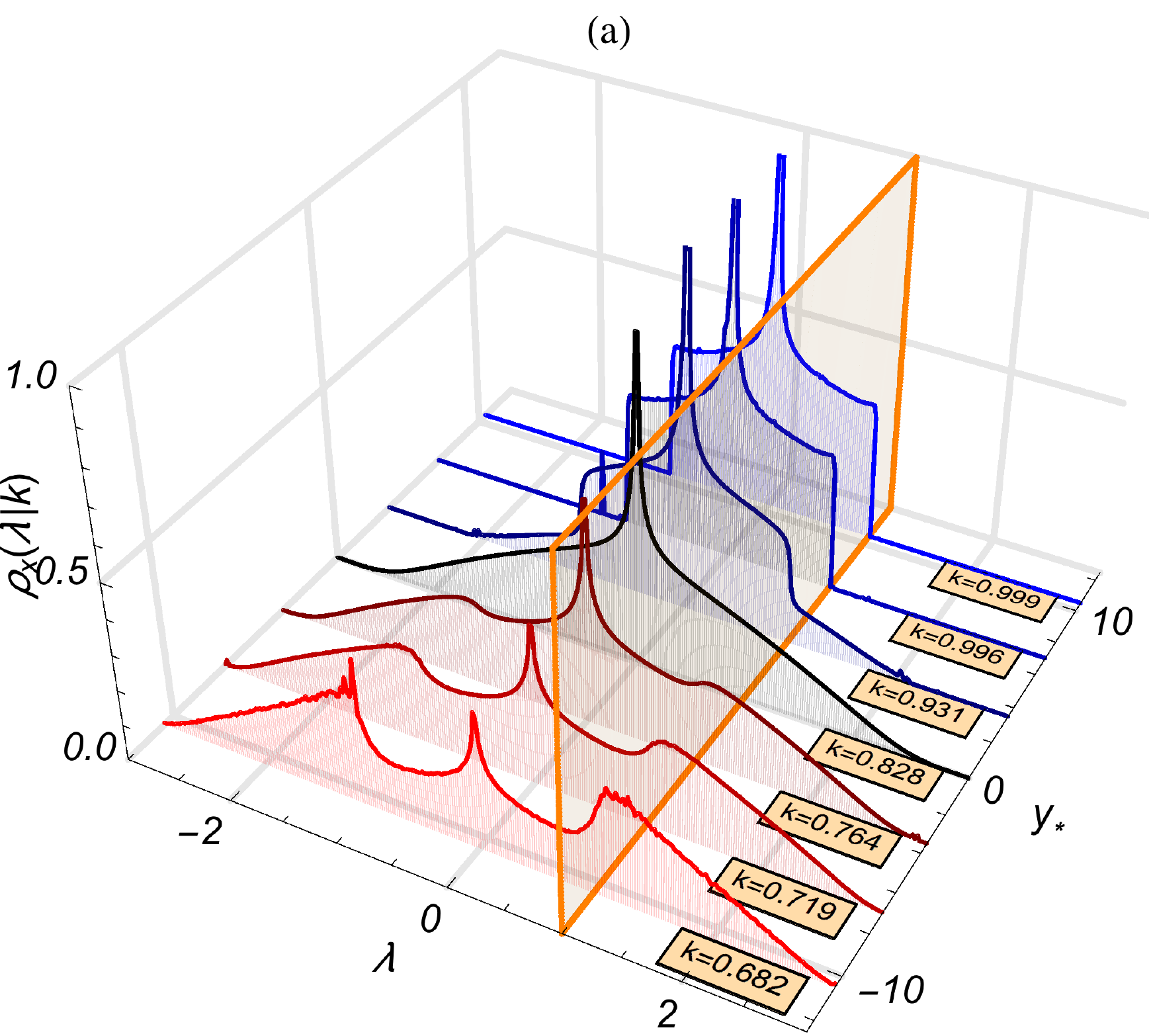} \\
\includegraphics[width=8.5cm,height=6cm]{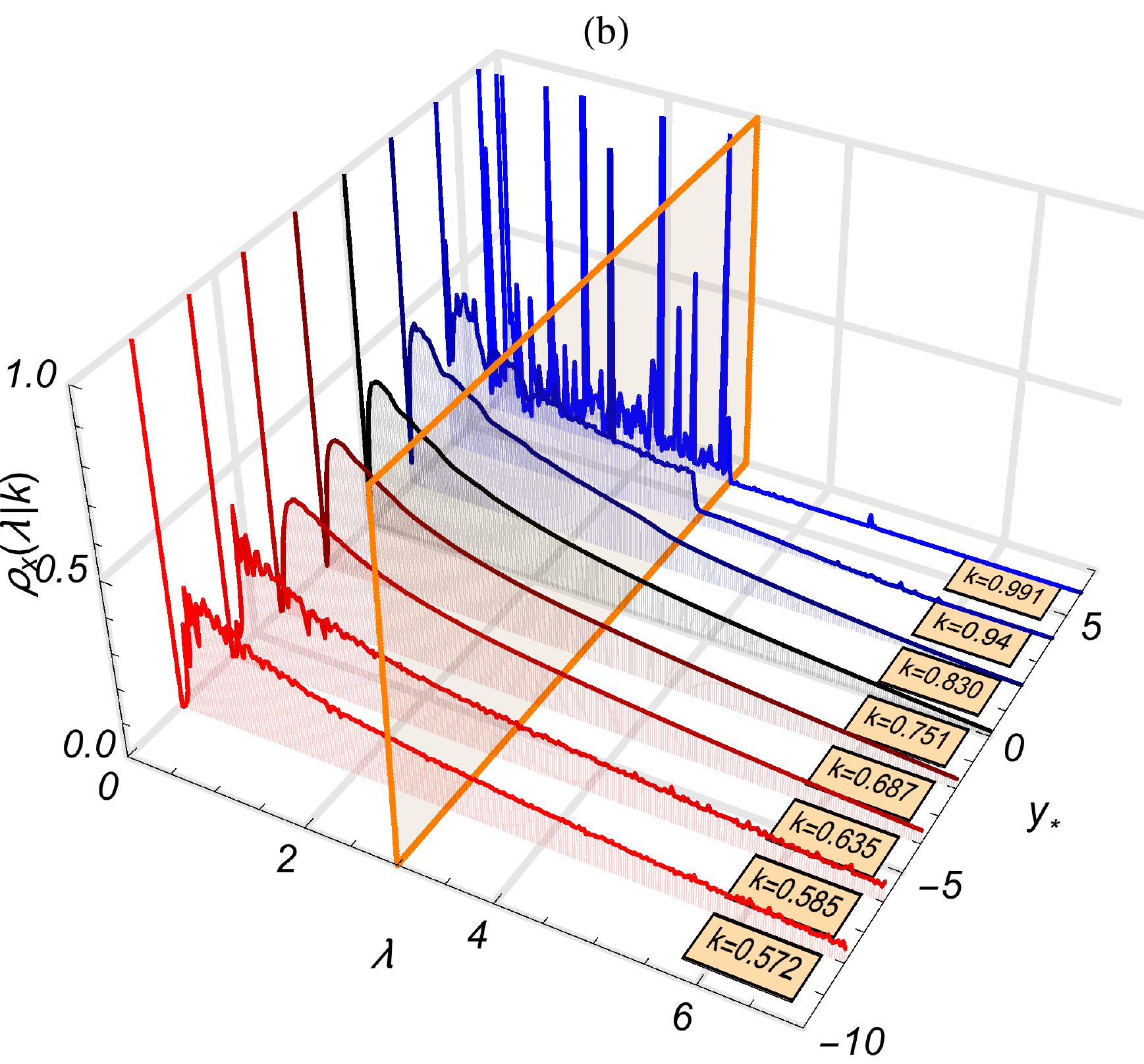}
\caption{
  Conditioned spectral density of Erd\"os-R\'enyi (ER) random graphs and sparse Wishart random matrices for fixed $x$ and several $y_{\star}$, derived from the numerical solution of our analytic equations. The unconditioned spectral density, where $k = k_{\text{typ}}(x)$, is obtained for $y=0$.
The values of $k$, indicated on the graphs, denote the fraction of eigenvalues smaller than $x$.
  (a) ER random graphs with average connectivity $c=5$ and Gaussian entries with zero mean and variance $1/c$. The wall is located at $x=1$. (b) Sparse Wishart random matrices with parameters $c= 2$ and $\alpha=2$. The wall is located at $x=3$.
}
\label{fig2}
\end{figure}

In Figure \ref{fig2}, we illustrate the main outcome of our theory for the two ensembles of SRMs.
Figure \ref{fig2} shows the CSD of ER random graphs with $c=5$ and of the sparse Wishart ensemble with
parameters $(\alpha,c) = (2,2)$. For each random matrix ensemble, we have chosen a different value of $x$ and
several values of $k$. In striking contrast to invariant random matrix
ensembles \cite{Majumdar2009,Vivo,Majumdar3}, the CSD has a compact support, there is no accumulation of
eigenvalues around $x$, and the function $\rho_x(\lambda|k)$ does not experience any type of sudden transition when $k = k_{\text{typ}}(x)$. All these features, which seem to be universal within the realm of sparse random matrices, are due to the weak or absent repulsion between the eigenvalues. In fact, the spectrum of sparse random matrices often contains localized eigenstates \cite{Kuhn2008,Metz2010,Slanina2012}, where the spacing between adjacent eigenvalues typically follows a Poisson distribution \cite{Evangelou1,Jackson,Slanina2012} and the eigenvalues can be arbitrarily close to each other.

\begin{figure}
  \centering
    \includegraphics[width=7.5cm,height=5cm]{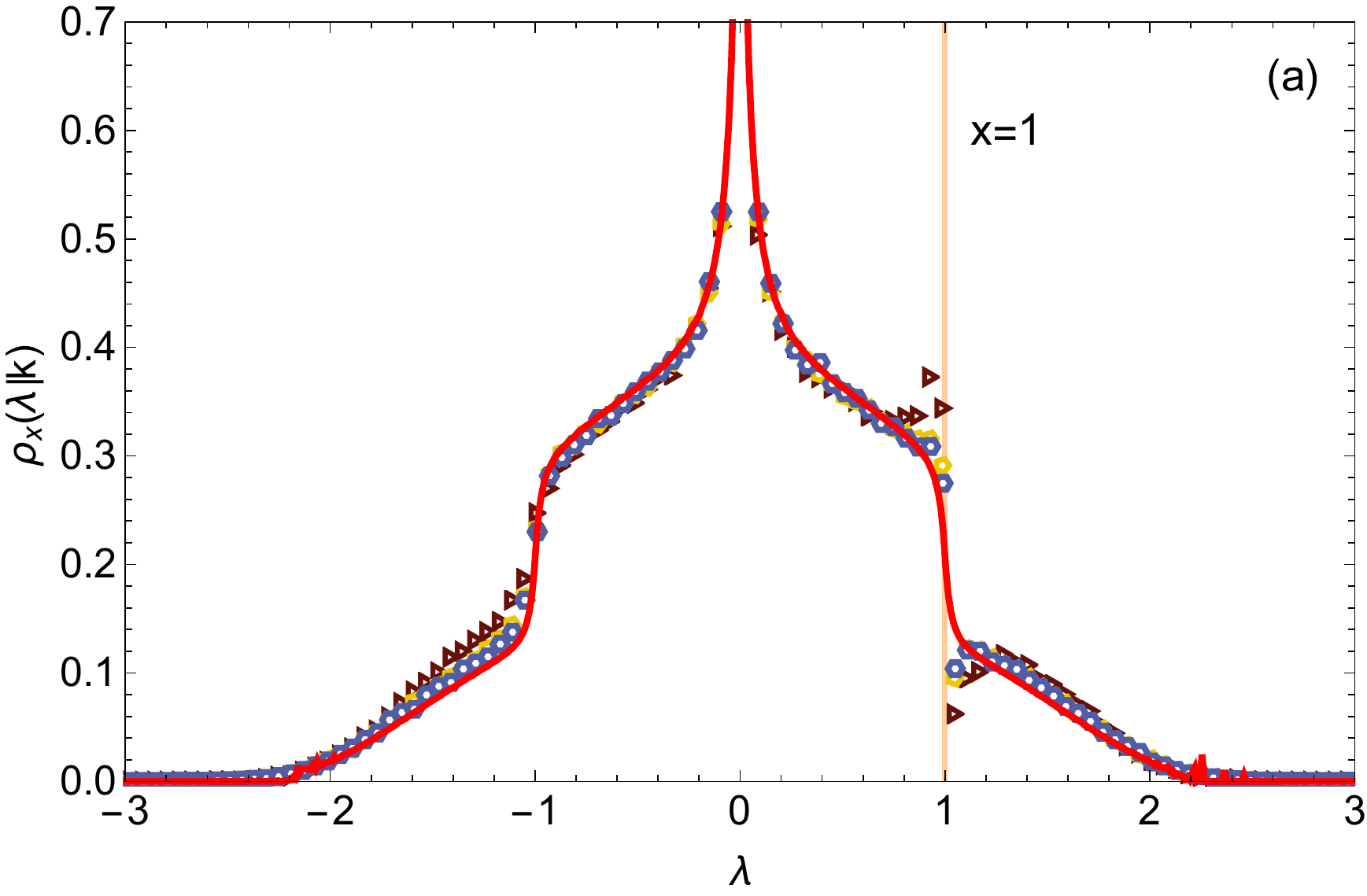} \\
\includegraphics[width=7.5cm,height=5cm]{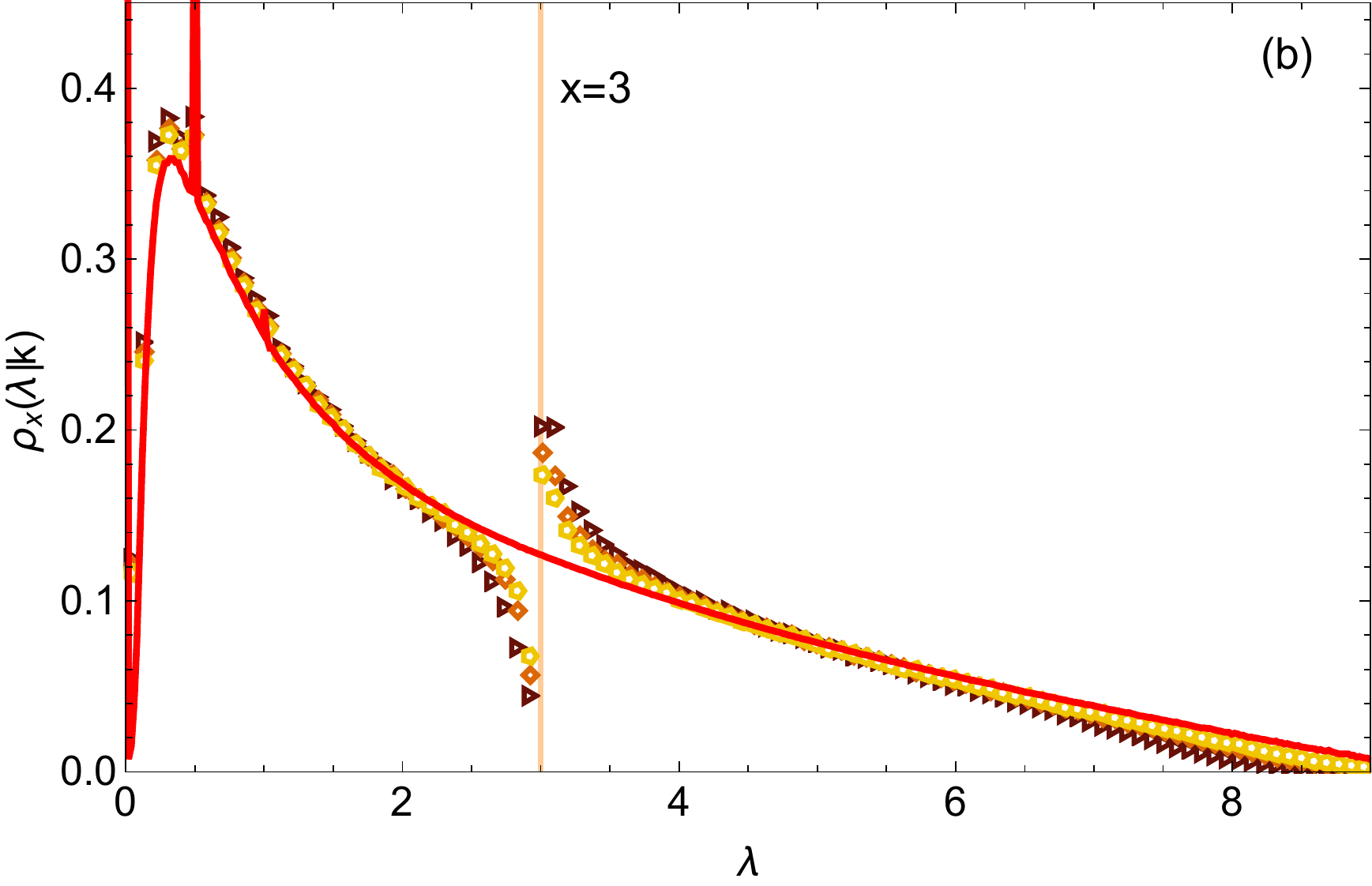}
   \caption{Comparison between the theoretical results (red curve) and numerical diagonalization (symbols) of finite random matrices for the  conditioned spectral density of  Erd\"os-R\'enyi (ER)  random graphs and sparse Wishart random matrices. The symbols result from the numerical diagonalization of  random matrices of dimensions  $N=25$ (brown triangles), $N=50$ (orange squares), $N=75$ (yellow pentagons), and $N=100$ (blue hexagons).  (a) ER random graphs with average connectivity $c=5$ and Gaussian entries with zero mean and variance $1/c$. The fraction of  eigenvalues smaller than $x=1$ is given by $k=0.92$, which is {\it larger} than its typical value $k_{\text{typ}}(x)=0.83$.  (b) Sparse Wishart random matrices with parameters $c= 2$ and $\alpha=2$.  The fraction of eigenvalues smaller than $x=3$ is given by $k=0.64$, which is {\it smaller} than its typical value $k_{\text{typ}}(x)=0.75$.} 
\label{fig1}
\end{figure}

We also note that the spectrum of infinitely large ER random graphs must be always symmetric. Strictly in the limit  $N \rightarrow \infty$, such symmetry requirement is individually fulfilled by each member of the graph ensemble. This implies that, mathematically, the constrained graph ensemble of Fig. \ref{fig2}(a) must be empty for $k<1/2$, signaling a transition for some  critical value $k_{c}$, which can be equal or larger than $1/2$. Because of this proximity to a transition, the time required to find
$y_{\star}$ slows down and it  becomes cumbersome to decrease $k$ towards $1/2$. In the case of the sparse Wishart ensemble, Figure \ref{fig2}(b) shows that, as $k \rightarrow 1^{-}$, the constraint is obeyed by a significant increase of the mass corresponding to the discrete part of the spectrum, while its continuous part decreases accordingly. It is difficult, though, to ascertain, either analytically or numerically, whether there is a critical value of $k$ above which the continuous part of $\rho_x(\lambda|k)$ vanishes.

Finally, let us compare  our theory with numerical diagonalization of finite matrices. Figure \ref{fig1} illustrates the CSD for
ER random graphs with $c=5$ and for sparse Wishart random matrices with parameters $(\alpha,c)=(2,2)$. The nonzero entries in the case of ER random graphs are drawn from $P_J(J)= \left(2\pi/c\right)^{-1/2} e^{- c J^2 /2 }$. Since we are exploring rare events and  advanced biased sampling methods are usually tailored for invariant random matrices \cite{Saito2010}, numerical diagonalization only enables us to extract the CSD for rather small values of $N$. Thus, finite size effects in figure \ref{fig1} are remarkable, showing that for finite $N$ the eigenvalues accumulate on both sides of the wall located at $x$. However, this effect is suppressed as the matrix dimension becomes larger and larger, with  numerical diagonalization results consistently approaching our theoretical predictions, valid strictly for $N \rightarrow \infty$. Overall, the agreement between our theory and numerical simulations is remarkable, which corroborates the exactness of our analytic equations. More than that, the method yields results beyond numerical diagonalization: it allows to determine the CSD in domains that are not accessible through diagonalization.

In this Letter we have put forward a powerful analytic approach to evaluate, in the limit $N \rightarrow \infty$, the spectral density of random matrices conditioned to have a fixed fraction $k$ of eigenvalues smaller than a threshold $x$. The present theory can be applied to the broad and scarcely explored class of non-invariant random matrices, for which the traditional Coulomb fluid approach is unworkable. We have shown how our approach can be used to unveil universal features of the conditioned spectral density in sparse random matrix ensembles. For both the adjacency matrix of Erd\"os-R\'enyi random graphs and sparse Wishart random matrices, the constrained density has a compact support, the eigenvalues do not accumulate at the position $x$, and there is no abrupt transition when $k=k_{\text{typ}}(x)$. On the other hand, our work suggests other types of transitions in the constrained density, such as the vanishing of the continuous part of the spectrum. All these features seem to follow from the weak or absent repulsive interaction between the eigenvalues, which is the driving force behind the properties of the constrained density in invariant random matrices.

The present work can be further developed in various directions, which extend beyond the scope of random matrices. Our analytic method allows to determine the spectral density of a diversity of random graph ensembles with structural constraints, shedding light in the relationship between the spectrum and the structure of complex networks \cite{Dorog}. The present approach can be also employed to explore equilibrium and non-equilibrium properties of rare samples of disordered systems, such as spin-glasses and combinatorial optimization  problems \cite{MezardBook}. The behavior of phase transitions in constrained ensembles of disordered systems is a general exciting problem, which is now accessible for analytical scrutiny.

\begin{acknowledgments}
FLM and IPC thank London Mathematical Laboratory for financial support. FLM acknowledges financial support from
CNPq (Edital Universal 406116/2016-4). This work has been funded by the program UNAM-DGAPA-PAPIIT IA101815. The authors thank the support of DGTIC for the use of the HP Cluster Platform 3000SL, codename Miztli, under the Mega-project LANCAD-UNAM-DGTIC-333. 
\end{acknowledgments}

\bibliography{biblio}

\end{document}